\documentclass[10pt, epsfig]{article}                              
\usepackage{epsfig}          
\usepackage{amssymb}     
\usepackage{amsfonts}    
\newskip\humongous \humongous=0pt plus 1000pt minus 1000pt

\newif\ifdtup

\catcode`\@=11
 
\@addtoreset{equation}{section}
\def\theequation{\thesection\arabic{equation}}
 
\def\@normalsize{\@setsize\normalsize{15pt}\xiipt\@xiipt
\abovedisplayskip 14pt plus3pt minus3pt%
\belowdisplayskip \abovedisplayskip
\abovedisplayshortskip \z@ plus3pt%
\belowdisplayshortskip 7pt plus3.5pt minus0pt}
 
\def\small{\@setsize\small{13.6pt}\xipt\@xipt
\abovedisplayskip 13pt plus3pt minus3pt%
\belowdisplayskip \abovedisplayskip
\abovedisplayshortskip \z@ plus3pt%
\belowdisplayshortskip 7pt plus3.5pt minus0pt
\def\@listi{\parsep 4.5pt plus 2pt minus 1pt
     \itemsep \parsep
     \topsep 9pt plus 3pt minus 3pt}}
 
\relax

\catcode`@=12
 
\catcode`\@=11
 
\def\section{\@startsection{section}{1}{\z@}{3.5ex plus 1ex minus
   .2ex}{2.3ex plus .2ex}{\large\bf}}
 
\def\thesection{\arabic{section}.}

\def\appendix{\setcounter{section}{0}
 \def\thesection{Appendix \Alph{section}:}
 \def\theequation{\Alph{section}.\arabic{equation}}}
 
\def\YGrule{0.4}   
\def\YGbox{6.5}    
\def\SymBoxes#1#2#3#4{\newdimen\un@t \un@t#3%
\raisebox{#1}{\rule{#2\un@t}{#4}\hskip-#2\un@t
\@tempdimb\un@t \advance\@tempdimb by-#4\@tempcntb#2\relax%
\@whilenum{\@tempcntb>0}\do{
\rule{#4}{\un@t}\hskip\@tempdimb \advance\@tempcntb by\m@ne}%
\hskip-#2\un@t \rule[\un@t]{#2\un@t}{#4}%
\rule[\un@t]{#4}{#4}\hskip-#4
\rule{#4}{\un@t}}\hskip-#4}                
\def\Young{\@ifnextchar[{\@Young}{\@Young[0]}}
\def\@Young[#1]#2{\newdimen\YG@unit \YG@unit\YGbox pt%
\newdimen\h@ight \h@ight#1\YG@unit \@tempcnta-1\relax
\@tfor\c@ount:=#2\do{\advance\@tempcnta by\@ne}
\@tempdima\@tempcnta\YG@unit%
\advance\h@ight by\@tempdima\relax     
\@tfor\c@ount:=#2\do{\SymBoxes{\h@ight}{\c@ount}{\YG@unit}{\YGrule pt}%
\@tempdima-\c@ount\YG@unit \hskip\@tempdima%
\advance \h@ight by -\YG@unit}         
\@tempdima\YG@unit \multiply\@tempdima by\@car#2\@nil %
\hskip\@tempdima}                      
\def\YoungTab{\@ifnextchar[{\@YoungIdx}{\@YoungIdx[0]}}
\def\@YoungIdx[#1]{\@ifnextchar[{\@iYoungIdx[#1]}{\@iYoungIdx[#1][\@empty]}}
\def\@iYoungIdx[#1][#2]#3{%
\newdimen\YG@unit \YG@unit\YGbox pt\newdimen\YG@rule \YG@rule \YGrule pt
\newcount\c@ount \c@ount\z@ \newdimen\skip@wd \unitlength\@ne pt
\newdimen\h@ight \h@ight#1\YG@unit \@tempcnta\m@ne\relax
\@tfor\d@um:=#3\do{\advance\@tempcnta by\@ne}
\@tempdima\@tempcnta\YG@unit%
\advance\h@ight by\@tempdima\relax
\@tfor\@idxlist:=#3\do{
\@tempcnta\z@\hskip.5\YG@rule\relax 
\@for\@idx:=\@idxlist\do{
\raisebox{\h@ight}{\makebox(\YGbox,\YGbox){#2$\@idx$}}
\advance\@tempcnta by\@ne}\hskip-.5\YG@rule%
\@tempdima-\@tempcnta\YG@unit \hskip\@tempdima%
\ifnum\c@ount=\z@ \skip@wd-\@tempdima\fi \relax
\SymBoxes{\h@ight}{\@tempcnta}{\YG@unit}{\YG@rule}%
\hskip\@tempdima \advance\h@ight by -\YG@unit
\advance\c@ount by\@ne}
\hskip\skip@wd}                      
   
\def\YGrule{0.4}   
\def\YGbox{6.5}    
\def\SymBoxes#1#2#3#4{\newdimen\un@t \un@t#3%
\raisebox{#1}{\rule{#2\un@t}{#4}\hskip-#2\un@t
\@tempdimb\un@t \advance\@tempdimb by-#4\@tempcntb#2\relax%
\@whilenum{\@tempcntb>0}\do{
\rule{#4}{\un@t}\hskip\@tempdimb \advance\@tempcntb by\m@ne}%
\hskip-#2\un@t \rule[\un@t]{#2\un@t}{#4}%
\rule[\un@t]{#4}{#4}\hskip-#4
\rule{#4}{\un@t}}\hskip-#4}                
\def\Young{\@ifnextchar[{\@Young}{\@Young[0]}}
\def\@Young[#1]#2{\newdimen\YG@unit \YG@unit\YGbox pt%
\newdimen\h@ight \h@ight#1\YG@unit \@tempcnta-1\relax
\@tfor\c@ount:=#2\do{\advance\@tempcnta by\@ne}
\@tempdima\@tempcnta\YG@unit%
\advance\h@ight by\@tempdima\relax     
\@tfor\c@ount:=#2\do{\SymBoxes{\h@ight}{\c@ount}{\YG@unit}{\YGrule pt}%
\@tempdima-\c@ount\YG@unit \hskip\@tempdima%
\advance \h@ight by -\YG@unit}         
\@tempdima\YG@unit \multiply\@tempdima by\@car#2\@nil %
\hskip\@tempdima}                      
\def\YoungTab{\@ifnextchar[{\@YoungIdx}{\@YoungIdx[0]}}
\def\@YoungIdx[#1]{\@ifnextchar[{\@iYoungIdx[#1]}{\@iYoungIdx[#1][\@empty]}}
\def\@iYoungIdx[#1][#2]#3{%
\newdimen\YG@unit \YG@unit\YGbox pt\newdimen\YG@rule \YG@rule \YGrule pt
\newcount\c@ount \c@ount\z@ \newdimen\skip@wd \unitlength\@ne pt
\newdimen\h@ight \h@ight#1\YG@unit \@tempcnta\m@ne\relax
\@tfor\d@um:=#3\do{\advance\@tempcnta by\@ne}
\@tempdima\@tempcnta\YG@unit%
\advance\h@ight by\@tempdima\relax
\@tfor\@idxlist:=#3\do{
\@tempcnta\z@\hskip.5\YG@rule\relax 
\@for\@idx:=\@idxlist\do{
\raisebox{\h@ight}{\makebox(\YGbox,\YGbox){#2$\@idx$}}
\advance\@tempcnta by\@ne}\hskip-.5\YG@rule%
\@tempdima-\@tempcnta\YG@unit \hskip\@tempdima%
\ifnum\c@ount=\z@ \skip@wd-\@tempdima\fi \relax
\SymBoxes{\h@ight}{\@tempcnta}{\YG@unit}{\YG@rule}%
\hskip\@tempdima \advance\h@ight by -\YG@unit
\advance\c@ount by\@ne}
\hskip\skip@wd}                      

\begin{document}

\newcommand{\beq}{\begin{equation}}
\newcommand{\eeq}{\end{equation}}
\newcommand{\bea}{\begin{eqnarray}}
\newcommand{\eea}{\end{eqnarray}}
\newcommand{\beas}{\begin{eqnarray*}}
\newcommand{\eeas}{\end{eqnarray*}}
\newcommand{\defi}{\stackrel{\rm def}{=}}
\newcommand{\non}{\nonumber}   
\newcommand{\bquo}{\begin{quote}}
\newcommand{\enqu}{\end{quote}}
\def\de{\partial}
\def\Tr{ \hbox{\rm Tr}}
\def\const{\hbox {\rm const.}}
\def\o{\over}
\def\im{\hbox{\rm Im}}
\def\re{\hbox{\rm Re}}  
\def\bra{\langle}\def\ket{\rangle}
\def\Arg{\hbox {\rm Arg}}
\def\Re{\hbox {\rm Re}}
\def\Im{\hbox {\rm Im}}
\def\diag{\hbox{\rm diag}}
\def\longvert{{\rule[-2mm]{0.1mm}{7mm}}\,}

\begin{titlepage}
{\hfill     IFUP-TH/2002-25} 
\bigskip
\bigskip
\bigskip
\bigskip

\begin{center}
{\Large  {\bf 
  NON-ABELIAN MAGNETIC MONOPOLES AND DYNAMICS OF  CONFINEMENT
} }
\end{center}

\vspace{4 em}

\begin{center}
{\large {\bf Stefano BOLOGNESI}}  

\vspace{1em}

{\it
Scuola Normale Superiore - Pisa  \\
Istituto Nazionale di Fisica Nucleare -- Sezione di Pisa  }
\\
{\it Piazza dei Cavalieri 7,   Pisa, Italy  }

\vskip 1cm 
 
 {\large  {\bf   Kenichi  KONISHI}}

\vspace{1em}

{\it
Dipartimento di Fisica   ``E. Fermi"  -- Universit\`a di Pisa\\
Istituto Nazionale di Fisica Nucleare -- Sezione di Pisa}
\\
{\it Via Buonarroti, 2,   Ed. C, 56127  Pisa, Italy 
\\    konishi@df.unipi.it     }
\end{center}

\vspace{7em}
\noindent  
{\bf Abstract:}

 {  Magnetic monopoles  having  non-Abelian charges have been found  recently to play a
crucial role in the infrared  in a class of  supersymmetric gauge theories.  We
argue that these ``dual quarks"    can  naturally be identified with   the   non-Abelian  magnetic monopoles of the type first
discussed by Goddard,  Nuyts and Olive.   Our argument is based on a few   simple observations as regards to  their charge structure,  flavor
quantum numbers,  and some general properties of electromagnetic duality.  
}

\vfill
 
\begin{flushright}
July  2002
\end{flushright}
\end{titlepage}

\bigskip

\section{Introduction}

With the advent of the exact Seiberg-Witten solution in    ${\cal N}=2$  supersymmetric gauge theories \cite{SW1,SW2,curves},  our
understanding  of nonperturbative dynamics of non-Abelian gauge theories received a significant boost. In particular,  the behaviors of
topologically nontrivial excitations such as magnetic monopoles or dyons,   which were  earlier known mainly from  semi-classical 
analyses,   can now be studied exactly in wide  classes of models.  
For instance,   many  examples of  strongly interacting  systems have been found  where the 't Hooft-Mandelstam mechanism of confinement
\cite{TM} is indeed  realized.

However, a series of studies in a wider class of softly broken ${\cal N}=2$  supersymmetric gauge
theories have shown \cite{ArPlSei,CKM} that   actually   much richer varieties of dynamical  possibilities exist.      Confining vacua in
$SU(n_c)$,
$USp(2n_c)$  or 
$SO(n_c)$  gauge theories with softly broken ${\cal N}=2$  supersymmetry,     with various  number of flavors   $n_f  <  2 n_c, \,\,  2 n_c+2,\,\,
n_c -2$, respectively,  have been found  to fall, roughly speaking,  into  the following three categories \cite{CKM}:   
 \begin{description}
\item{(i)}   In some of the vacua  (the $r=0$  or $r=1$  vacua of $SU(n_c)$ theories;  also confining vacua of all flavorless cases),  the gauge group of the
low-energy dual theory is  the maximal    Abelian subgroup  $U(1)^R$, where  $R$ is the rank of the original gauge group;   confinement  is
described by 't Hooft-Mandelstam mechanism; 
\item{(ii)}   In the general  $r$ vacua  ($2 \le r  < {n_f \o 2 } $)  of the  $SU(n_c)$ theory,   the effective low-energy theory is a   non-Abelian
$SU(r) \times  U(1)^{n-r}$  gauge  theory;   among the massless degrees of freedom are the magnetic monopoles in the fundamental
representation of  dual
$SU(r)
$  gauge group.  Their condensation, together with that of Abelian monopoles of the  $U(1)^{n-r}$ 
factors, describes the confinement of the original electric theory as a generalized dual Meissner effect.  The vacua in the same universality
classes  appear in 
$USp(2n_c)$ and 
$SO(n_c)$ theories  with nonzero bare hypermultiplet  (quark) masses; 
\item{(iii)}  In the $r= { n_f \o 2}$   vacua of  $SU(n_c)$ theory,  as well as in  {\it all}  of the confining vacua of $USp(2n_c)$ and  $SO(n_c)$ 
theories with  vanishing bare quark masses \footnote{ There are  exceptions to this rule  for small values of $n_f$ and $n_c$, 
e.g., 
 $USp(2)= SU(2)$ case.  See the
footnote 18  of 
\cite{CKM}.},  the low-energy degrees of freedom involve relatively non-local states: the effective   theory is a deformed superconformal
theory,
i.e., near an infrared fixed-point. There is no local Lagrangian description of  such  vacua. 

\end{description} 

The aim of this paper is to  argue  that the  ``dual quarks"   appearing in the $r$-vacua  ($r \ge 2$)  of the softly broken ${\cal N}=2$ 
$SU(n_c)$ theories,   (ii) above,   are nothing but    the   non-Abelian  magnetic
monopoles of the type first discussed by Goddard,  Nuyts and Olive \cite{GNO} and studied extensively  by E. Weinberg \cite{EW}.   Our
argument is based on the simple observations as regards to  their charges,  flavor quantum numbers,  and some general properties of
electromagnetic duality.

\section{ Semiclassical Non-Abelian Monopoles \label{sec:semic} }

Consider     a  gauge theory where the gauge group $G$ is  broken,
$$   G   \,\,\,{\stackrel {\bra \phi \ket    \ne 0} {\Longrightarrow}}     \,\,\, H  $$ 
where  the unbroken group  $H$  is  in general  non-Abelian.   In order to have a nontrivial finite-energy configuration,  the scalar  fields and
gauge field must behave asymptotically as  
\beq    {\cal D} \phi    \,\,\,{\stackrel {r \to   \infty  } {\longrightarrow}}   \,\,\,0       \quad    \Rightarrow   \quad 
\phi \sim   U \cdot  \bra \phi \ket  \cdot U^{-1}, \qquad    A_i^a  \sim  U \cdot {\de_i  }   U^{\dagger}   \to     
\epsilon_{aij}  { r_j 
\o     r^3}   G(r),        \eeq
representing nontrivial elements of   $  \Pi_2(G/H)  = \Pi_1(H).$
The  function   $G(r)$ can be chosen as
\beq    G(r)  =      \beta_i   T_i,   \qquad   T_i  \in  {\hbox {\rm  Cartan Subalgebra of}} \,\,  H.  
\eeq
Topological quantization leads to the result \cite{GNO}  that   the  ``charges"  $ \beta_i $   take values which are  
weight vectors of the group   ${\tilde H} $ where  ${\tilde H}  =    {\hbox {\rm  dual   of}} \,\, H. $
The dual of  a group (whose  roots vectors are $\alpha$'s)   is defined by  the root vectors  which span  the dual lattice,
i.e.,    ${\tilde \alpha }= \alpha /  \alpha^2.$
Examples of pairs of the duals  are: 

{\large 
\begin{table}[h]   
\begin{center}
\begin{tabular}{c  c   c}
\hline  
$SU(N)/Z_N       $        &   $\Leftrightarrow$                 &    $SU(N)     $          \\
  $ SO(2N)  $     &   $\Leftrightarrow$    &   $SO(2N) $       \\  
  $ SO(2N+1)  $     &   $\Leftrightarrow$     &   $USp(2N) $       \\ \hline
\end{tabular}
\label{tabtheta}
\caption{Some examples of dual pairs of groups}
\end{center}
\end{table}}

To be concrete we  consider a general (supersymmetric or non supersymmetric)  $SU(n_c)$  gauge theory with an appropriate set of scalar
fields  in the adjoint representation.  As will be mentioned at the end, our analysis applies equally well to other gauge groups. 
We assume that
the minimum of the potential is such that the gauge group  is broken spontaneously as 
\beq     SU(n_c) \to    SU(r) \times   U(1)^{n_c -r}. 
\label{breaking}\eeq
For instance the  VEV  of a scalar can be taken in the diagonal form
\beq   \bra \phi\ket =
 \pmatrix{  v_0  {\bf 1}_{r\times r}   & {\bf 0 } &  \ldots &   {\bf 0}   \cr  {\bf 0 }  & v_{r+1}    & \ldots   & 0   \cr  \vdots  &\vdots& \ddots
&
\vdots
\cr {\bf 0}    & 0   &  \ldots  & v_{n_c}     }, \qquad   r  \, v_0  +  \sum_{j=r+1}^{n_c}     v_j =0,
\label{simple} \eeq
where $v_i$'s are all different.  
Let us write the asymptotic Higgs field more compactly as
\beq     \phi_0  =  {\mathbf h} \cdot  {\mathbf H},  
\eeq
where   the $n_c-1$ rank vector  $ {\mathbf h}$  describes the scalar VEV,   while $ {\mathbf H}$  represents the generators in the Cartan subalgebra 
of $SU(n_c)$.  If $ {\mathbf h}$ had non-zero inner products with all of the root vectors  of $SU(n_c)$  then the gauge group would be  maximally broken to
$U(1)^{n_c-1}$ group  and Abelian monopoles having respective $U(1)$ charges would appear.  We have nothing to add about such a
system.   

    Here we consider the case in which  $ {\mathbf h}$  is  orthogonal  to  the root vectors of a $SU(r)$ subgroup.  
The simplest way to detect the presence of the non-Abelian monopoles is to consider various  $SU(2)$  subgroups generated by
\beq     t_1=  { 1\o \sqrt{ 2 { \mathbf \alpha}^2}  }  (  E_{{ \mathbf \alpha}}   +   E_{-{ \mathbf \alpha}}     ); \qquad  
 t_2=  - { i \o \sqrt{ 2 {\mathbf \alpha}^2  }    }(  E_{{ \mathbf \alpha}}   -     E_{-{ \mathbf \alpha}} ); \qquad  
t_3=   {\mathbf \alpha}^{*} \cdot  {\mathbf H}, 
\eeq
where ${\mathbf  \alpha}$    is a root vector associated with     broken generators    $ E_{\pm{ \mathbf \alpha}}$  and        $ {\mathbf
\alpha}^{*}
\equiv  {\mathbf
\alpha}  /  {\alpha^2}. $  In particular we consider those  $SU(2)$ groups which  do not commute with the unbroken subgroup
$SU(r)$.       In the notation of Eq.(\ref{simple}) these  
    correspond to    $SU(2)$  subgroups   acting in the $[i-k]$  subspaces,  where  $i=1,2,\ldots, r,\,\,$  and   
$k=r+1, r+2,
\ldots  n_c$.  
 The symmetry breaking  (\ref{breaking}) induces the Higgs mechanism in such an $SU(2)$  subgroup, 
\beq    SU(2)  \Longrightarrow  U(1).
\eeq
By  embedding the known   't Hooft-Polyakov monopole \cite{TP}     lying in  this subgroup, and 
 adding a constant term for $\phi$   so that it behaves  correctly  asymptotically,  one easily constructs a  solution of the $SU(n_c)$ 
equation of motion (see E. Weinberg \cite{EW}):
\beq   A_i({\bf r})  =  A_i^a({\bf r},  {\bf h} \cdot {\mathbf \alpha}) \, t_a;  \qquad \phi({\bf r}) =     \chi^a({\bf r},  {\bf h} \cdot {\mathbf
\alpha})
\, t_a   +  (  {\bf h}   -   ({\bf h}
\cdot {\mathbf \alpha})    {
\mathbf
\alpha}^{*}  )
\cdot {\mathbf H},
\label{NAmonop}\eeq
where 
\beq    A_i^a({\bf r}) =  \epsilon_{aij}  { r^j \o r^2}  A(r); \qquad   \chi^a({\bf r}) =  { r^a \o r} \chi(r), \qquad    \chi(\infty)=
  {\bf h} \cdot {\mathbf \alpha}
\eeq
is  the standard 't Hooft-Polyakov solution.  Note that $\phi({\bf r}=(0,0,\infty) ) = \phi_0.$  

 The mass of this monopole for the minimum
magnetic charge is given by the standard formula (we assume the BPS situation) 
\beq     M=  { 4 \pi \o g}   {\bf h} \cdot {\mathbf \alpha} =   { 4 \pi \o g} | v_0 - v_k | .
\eeq
By an appropriate field redefinition $v_0$   can be always taken to be positive.  Also,   for generic, unequal  values of $v_i$, 
 it is possible, by using a Weyl
transformation,   to  take  the scalar VEV so that  
\beq   | v_0 - v_{r+1}  |<   | v_0 - v_{k} |, \qquad   k = r+2, r+3, \ldots,   n_c.
\eeq 
By considering various  $SU(2)$  subgroups acting on $[i, r+1]$ subspaces,   where   
$i=1,2,\ldots, r$,   we  find   that 
{\it there are precisely $r$  degenerate   solutions with the same mimimum   mass, \beq M= { 4
\pi
\o g} | v_0 - v_{r+1}|. \label{minimum} \eeq }      They are transformed to each other by the Weyl transformations.   By
construction these solutions carry also a unit (magnetic)  charge with respect to the   
$U_0(1)$ gauge group, which is generated by
\beq      Q_0 
 =   \pmatrix{ { 1\o r}   {\bf 1 }   & { 0 }    &  \ldots    &   \ldots    \cr  
                          { 0 }   &  -1   & 0  &  \ldots      \cr 
                          \vdots    & 0  & 0  &  \ldots    \cr  
                       \vdots    & 0  &  \ldots   &  \ddots   
        }  
\eeq

The system, furthermore,  has $n_c-r-1$  Abelian  monopoles,   each with the minimal charge  in 
\beq   Diag\,  Q_{\ell} =  [ \, {\underbrace  {0,0, \ldots,
0}_{\ell}},  1, -1, 0, \ldots, 0  \, ] , \qquad   r \le \ell \le n_c-1, 
\eeq  
and with mass
\beq     M_{\ell}=    { 4 \pi \o g} | v_{\ell} - v_{\ell+1}  |.
\eeq
For appropriate choice of the scalar vacuum expectation values (VEVS) (and arranging them appropriately by Weyl transformations)   there are
thus  an
$r$-plet of ``non-Abelian"  monopoles and
$n_c-r-1$   Abelian monopoles  with minimum charges and minimum masses  that are stable .

We are going to argue that the degenerate $r$-plet of monopoles can, under  appropriate conditions, 
emerge as a multiplet in the fundamental representation of the dual $SU(r)$   group.    Their quantum numbers, together with those of
Abelian monopoles,  would
appear  as in Table \ref{tabNA}    
  \begin{table}[h]
\begin{center}
\begin{tabular}{c  c   c c c  c c  }
 \\
  Monopoles    &    ${ {SU}}  (r)   $           &    ${ U_0(1) }     $      &    ${U_1(1) }     $     &    ${ U_2(1) }  $   &
$\ldots  $  &         ${ U_{ n_c-r-1} (1) }  $
\\    \hline
  $ q  $        &     $ {\underline r   }$     &      $1$     &  $0$ &   $0$ & \ldots   &  $0$    \\  \hline
 ${ e}_1  $        &     $ {\underline 1 }$     &
$0 $     &   $1$  &  $0$     & $\ldots $     &  $0$    \\  \hline
${ e}_2 $        &     $ {\underline 1  }$     &   $0 $     & $0 $  &  $1$   & $\ldots$    &  $0$  \\  \hline
$\vdots $        &     $ \vdots$     &      $\vdots $     &  $ \vdots $    &   $ \vdots $  &  $ \ddots $   &   $\vdots$    \\  \hline
${ e}_{n_c-r-1}    $        &     $ {\underline 1  }$     &      $0$     & $0$     &  $ 0$   &  $\ldots $     &  $1$   \\  \hline
\end{tabular}
\label{tabNA}
\caption{Stable magnetic monopoles of minimum masses and their charges  }
\end{center}
\end{table}

As an illustration of  the above construction,  consider the simplest nonotrivial case with $n_c=3, \,\, r=2$, i.e.,  an $ SU(3)$  theory
with symmetry breaking,
 \beq 
SU(3) {\stackrel {\bra \phi \ket } {\longrightarrow}}     SU(2) \times  U(1), \qquad   \bra \phi\ket = 
 \pmatrix{  v & 0& 0  \cr  0 & v & 0 \cr  0&0& -2v  }.
\eeq 
By considering a  broken $SU(2)$ subgroup   (``$U$"-spin), 
\beq   t^4= { 1\o 2}   \pmatrix{  0 & 0& 1  \cr  0 & 0 & 0 \cr  1  &0& 0   }; \quad  
t^5= { 1\o  2}   \pmatrix{  0 & 0& -i  \cr  0 & 0  & 0 \cr i &0& 0    }; \quad  
{ t^3 +  \sqrt3 t^8 \o 2} = { 1\o 2}   \pmatrix{  1  & 0  & 0  \cr  0 & 0 & 0 \cr  0&0& -1   }
\eeq
one finds a solution 
\bea   \phi ({\bf r})  &=&
   \left( \begin{array}{ccc}
     -\frac{1}{2}v&0&0\\
     0&v&0\\
     0&0&-\frac{1}{2}v\\
   \end{array} \right)
   +
3\,  v \Big( t_4,t_5,\frac{t_3} {2} + \frac{\sqrt{3} t_8}{2} \Big)
   \cdot \hat{r} \phi(r),   \non \\
  \vec{A}({\bf r}) &=&  \Big( t_4,t_5,\frac{t_3} {2} + \frac{\sqrt{3} t_8}{2} \Big)
   \wedge \hat{r} A(r),  \label{su3sol}\eea
where $\phi(r)$ and   $A(r)$ are 't Hooft's  functions  with  $\phi(\infty)  =1,\,$   $\phi(0)  =0,\,$  $A(\infty)   =-1/r.\,$   A second  solution
with the same energy can be constructed   by using another $SU(2)$  group (``$V"$-spin) acting in the $[2-3]$ subspace.   Together, they form
a doublet of the unbroken   (dual)
$SU(2)$ group.   They  carry a  unit  $t^8$   charge.    There are no other ``Abelian" monopoles.

Existence of the magnetic monopoles transforming  as in the fundamental representation of the {\it dual}  $SU(r)$  group,  might appear
to be in contradiction with   some   earlier  results  on  the non-existence of  dyons with non-Abelian charges  \cite{CDyons}, but it is not.  
Our monopoles carry nonzero charges with respect to $SU(r)$ as well as to one of the $U(1)$  factors,  but both refer to the   dual of the
original    subgroups. 
 The arguments  excluding   the possibility of  ``colored dyons"  (referring to the GUTs magnetic monopoles carrying the electric, color $SU(3)$
quantum numbers)  do not apply.  The crucial point     is that  the monopoles with non-Abelian charges (\ref{NAmonop}),   are  to  transform
linearly under the  dual   of the unbroken gauge group,  and not under the original unbroken subgroup.  

Note   that in the
presence of an unbroken, asymptotically-free gauge interactions,  the question of the  validity of semiclassical approximation sometimes
used  in these discussions,    is a subtle one.  For instance,  it is not justified  to smoothly approach the limit of non-maximal
breaking such as in (\ref{breaking}),  starting from the maximally  broken case,  by  letting some of the diagonal elements  of the scalar
VEVS to coincide.    Some of the Abelian  monopoles   would become apparently  massless  and spatially infinitely
extended  in such a limit;  any result involving these light states lies however  beyond the scope of the semi-classical approximation.   
 In fact,  it would  hardly 
make any sense to consider  the spontaneously broken  $SU(2){\stackrel {v} {\longrightarrow}}   U(1) $  theory and  to   take the
limit $v\to 0$    to attempt to find out what happens to the 't Hooft - Polyakov monopole, with  a semi-classical method!

For the same reasons   the system  characterized by the scalar  VEVS,  Eq.(\ref{simple}),  is never really semiclassical, even if 
\beq  v_0 \gg \Lambda;  \qquad  v_j \gg \Lambda, \quad  \forall j   \label{semicl}\eeq  
where   $\Lambda$ is the scale of the $SU(n_c)$  theory,     because the unbroken $SU(r)$  interactions become strong at long distances.    
Of course,  if (\ref{semicl}) were not satisfied,  semi-classical formulae like Eq.(\ref{minimum}) would  break down completely. 

However,    all is not lost.   As long as the breaking scales $v_0, v_j $ are much larger than the scale of the $SU(r)$  theory,     the properties of
the stable states such as their  multiplicity  and  charges   as  summarized  in Table \ref{tabNA}      should be  correct.  (Also, the mass
formulae should be approximately valid.)    They  represent the nonzero elements of
the homotopy group,
\beq    \Pi_2  (  { SU(n_c)  \o   SU(r) \times U(1)^{n_c-r}  } ) =  \Pi_1  (   U(1)^{n_c-r}    )   =   {\mathbf Z}^{n_c -r}.
\eeq
We assume that these properties are maintained as  Higgs VEVS  $v_0$  and $v_j $'s  are smoothly varied.  Their masses will vary, of course, 
in an unknown way.

The central question      is whether the unbroken $SU(r)$  gauge group is further {\it dynamically}   broken by the strong 
$SU(r)$  interactions themselves.  This depends on the
  system considered.  In
some model  the $SU(r)$ symmetry may be broken further, e.g., to $U(1)^{r-1}$.   In this case the "$r$-plet of  non-Abelian monopoles"
 simply
means the presence of $r$   approximately degenerate (as long as $v_0$, $\,v_j$'s  are all large  compared to the scale of
$SU(r)$ gauge theory)  monopoles \footnote{$N=2$ supersymmetric $SU(n_c)$ pure Yang-Mills theory is a good example of this types of
theory.  Indeed,  in the case of $SU(3)$  theory,  it can be seen that in an appropriate semi-classical domain  of  $u= \bra \Tr \Phi^2 \ket, $   
$v= \bra
\Tr  \Phi^3 \ket, $    such that   $a_1 = - a_2$,   there are two  approximately  degenerate monopoles.  See Eq.(6.4)  of Klemm
et. al. \cite{curves}.  }. 
 Only if $SU(r)$ is  not dynamically broken further do these particles behave truely as 
the fundamental multiplet  of the dual gauge group.

 What is most remarkable  in our opinion is   the fact that this second option seems to be realized in the 
$r$-vacua of the  softly broken ${\cal N}=2$  $SU(n_c)$ 
gauge  theory  \cite{CKM}.  
Massless  magnetic  particles with precisely the   properties  listed in  Table \ref{tabNA} 
   appear in the fully quantum-mechanical   low-energy effective  action there.  
We propose  that these are   non-Abelian monopoles of the type discussed above.

It is also  significant that, in the softly broken ${\cal N}=2$  $SU(n_c)$ theory,    the $r$ vacua  with a  magnetic $SU(r)$  gauge group 
occur only for   $r  \le   {n_f \o 2}$.      Before discussing this point further, we show in 
the next section  that if fermions  in the
fundamental  representation  of the  $SU(n_c)$  gauge group  are introduced in the theory,  each of them 
 possesses one   zero mode in the
background of  appropriate  non-Abelian  monopoles  of the form,  Eq.(\ref{NAmonop}).    It is  then possible that
 the non-Abelian monopoles   in the ${\underline  r}$ of $SU(r)$  transform    as a fundamental representation
 of the global  $SU(n_f)$ symmetry.   
 This explains the occurrence of non-Abelian monopoles carrying  $SU(n_f)$ quantum numbers   in the
$r$-vacua of the softly broken 
${\cal N}=2$  theory.

\section {Flavor Charges  of  Non-Abelian Nonopoles;  Confinement versus Dynamical Symmetry Breaking \label{JRM}   }

The mechanism with which the 't Hooft-Polyakov monopole acquires nontrivial flavor quantum numbers  is well-known \cite{JR,EW,Callias}. 
In order to show that an  analogous result holds for our non-Abelian monopoles, consider a fermion $\psi_{L,R}$   in the fundamental
representation (${\underline 3}$)  in the case of a $SU(3) $  theory  broken as  $SU(3)  \to SU(2) \times U(1)$  considered above.   
\( \psi_L \)
and 
\( \psi_R \)
can be decomposed into 
\( SU(2)_u \)  multiplets:
\beq \psi_L = \psi_{L(2)} \oplus \psi_{L(0)},  \qquad   \psi_R = \psi_{R(2)} \oplus \psi_{R(0)} \eeq
The zero-energy Dirac equations have  the form     \footnote{ The gamma matrices used  by  Jackiw and Rebbi,    
$\gamma_{JR}^0=\pmatrix{ {\bf 0} & -i {\bf 1} \cr i {\bf 1} & {\bf 0} };$  
$ \gamma_{JR}^i= \pmatrix{-i \sigma_i &  {\bf 0} \cr  {\bf 0} & i \sigma_i } $ and  
$ \gamma_{ch}^0=\pmatrix{{\bf 0}  &   {\bf 1} \cr {\bf 1} & {\bf 0} };\,\,$
$  \gamma_{ch}^i= \pmatrix{ {\bf 0}  & \sigma_i \cr  -\sigma_i  & {\bf 0}   } $   
    in the chiral representation used here,   are  related by
$S \gamma_{JR}^{\mu} S^{-1} =\gamma_{ch}^{\mu}$,  where    $  {S} =  {1\o2}  \pmatrix{ (1+i){\bf 1} & -(1+i){\bf 1}  
\cr (1-i){\bf 1} & (1-i){\bf
1} }$.  The Yukawa coupling has been  set to unity.  }
$$-\vec{\sigma} \cdot \vec{p} \psi_{L(2)} - e \vec{\sigma} \cdot
   (\vec{t} \wedge \hat{r}) A(r) \psi_{L(2)} - \frac{1} {2} v \, 
   \psi_{R(2)} + { 3  \, v } \,  \vec{t} \cdot \hat{r}
   \psi_{R(2)} \phi(r) = 0,  $$
\[ -\vec{\sigma} \cdot \vec{p} \psi_{L(0)} +  v \,  \psi_{R(0)} = 0,  \]
\[ \vec{\sigma} \cdot \vec{p} \psi_{R(2)} + e \vec{\sigma} \cdot
   (\vec{t} \wedge \hat{r}) A(r) \psi_{R(2)} - \frac{1} {2} v \, 
   \psi_{L(2)} + {3 v}\,  \vec{t} \cdot \hat{r}
   \psi_{L(2)} \phi(r) = 0,  \]
\beq  \vec{\sigma} \cdot \vec{p} \psi_{R(0)} +  v \psi_{L(0)} = 0,  \label{dirac}  \eeq   
where  
\( \vec{t} = \Big( t_4,t_5,\frac{t_3} {2} + \frac{\sqrt{3} t_8}{2} \Big) \).
These equations can be decoupled by using the combinations
\( \psi_+ = \psi_L + i\psi_R \),
\( \psi_- = \psi_L - i\psi_R \)
as ($m \equiv  { v \o 2}$) 
\[ -\vec{\sigma} \cdot \vec{p} \psi_{+(2)} - e \vec{\sigma} \cdot
   (\vec{t} \wedge \hat{r}) A(r) \psi_{+(2)} +im
   \psi_{+(2)} -i 3  \, v  \,  \vec{t} \cdot \hat{r}
   \psi_{+(2)} \phi(r) = 0, \]  
\beq -\vec{\sigma} \cdot \vec{p} \psi_{+(0)} -i  v \psi_{+(0)} = 0,   \label{eqsforpsip}  \eeq
and similarly for  $\psi_{-(2)} $  and   $\psi_{- (0)}$, with a minus sign in front of the corresponding terms 
with  $m$, $v$ or  $\phi(r)$.  
The singlet fermion obviously do not possess any  zero modes. 
As for 
\( \psi_{+(2)} \)
and   
\( \psi_{-(2)} \)  
Eq.(\ref{dirac}) are  formally the same as the equations for massive  fermions with mass,  $m \equiv  { v \o 2}$.
  According to the Callias index analysis  for massive fermions  there is one  normalizable zero mode if  
$  |\phi | -  |  m | >0; $   
 there are none otherwise  \cite{EW,Callias}.   In our normalization,   this condition reduces to
\beq 
\frac{3}{2} v >  { 1\o 2}  v  \label{unequal} \eeq
  which is obviously satisfied.   Therefore the doublet fermion possesses one zero mode. 

An analogous construction in the case of the breaking
$ SU(n_c) \to    SU(r) \times   U(1)^{n_c -r}, $
the above condition is replaced by 
\beq   \left|  { v_0  -   v_{r+1}  \o 2}    \right| >    \left|  { v_0   +     v_{r+1}  \o 2}    \right|.    \eeq
Note that for the breaking   $SU(n) \to  SU(n-1)\times U(1)$ such a condition is always satisfied; 
otherwise, only the monopoles with  VEVS satisfying   the above condition will give rise to fermion zero modes.

To construct explicitly the zero mode, we set 
\[\psi_{+(2)} = -i\sigma_{\alpha i}^{2}  \,  g_+(r) + (\vec{\sigma}\sigma^2)_{\alpha i} \hat{r} f_+(r), \]
\beq \psi_{-(2)} = -i\sigma_{\alpha i}^{2} \,  g_-(r) + (\vec{\sigma}\sigma^2)_{\alpha i} \, \hat{r} f_-(r) \eeq
in Eq.(\ref{eqsforpsip}) and analogous one for $ \psi_{-(2)}$.    The scalar functions   $g_{+}$, $f_{+}$
 satisfy     coupled linear differential equations,
\beq    g_{+}'(r) +G_+g_{+}(r)+ m f_{+}(r)=0, \qquad    f_{+}'(r) +F_+f_{+}(r)+ m g_{+}(r)=0,    \eeq
and similarly  $(g_{-}, \,  f_{-})$    with $(-)$ in front of the mass terms. 
The functions appearing in the coefficients are
$  G_+(r) = \frac{3}{2}v \phi(r)+ eA(r),\, $  
$ G_-(r) = -\frac{3}{2}v \phi(r)+ eA(r),\, $
$ F_+(r) = \frac{3}{2}v \phi(r)- eA(r)+\frac{2}{r},\, $
$ F_-(r) = -\frac{3}{2}v \phi(r)- eA(r)+\frac{2}{r}.  $
By eliminating  $g_+$ in favor of  $f_+$  (or {\it vice versa}) we get a second order differential equation involving 
 $f_+$ only    (or   $g_+$ only).   From the asymptotic behavior of the solutions of  these  equations we find that if  the condition (\ref{unequal})
is satisfied,  there is one solution  $(g_{+}, \,  f_{+})$   regular at the origin,       which is normalizable.  Both solutions  $(g_{-}$, $f_{-})$ are 
 instead   non-normalizable.

In the case of a $n_f$-flavored model,    then, each  fermion   has one zero mode,  in the background of  each  monopole. 
Denoting   the quantized fermion field as   $\psi_i  =    b_i   \psi^{(0)} +  \ldots, \,\,\, $   ($i=1,2,\ldots, n_f$),   where $b_i$'s are the zero mode
annihilation operator, the    standard procedure   yields  the possible  monopole  multiplets,   
\beq   |k\ket,  \qquad     b_i^{\dagger}|k\ket,       \qquad      b_i^{\dagger}  b_j^{\dagger}|k \ket, \qquad   \ldots, \qquad   (k=1,2,\ldots, r;
\,\,i,j=1,2,\ldots,  n_f) 
\label{possib}  \eeq 
  which belong to  various  antisymmetric irreducible representations of the flavor $SU(n_f)$  group  (${\bf
1},\,$  $\Young{1},\,$  $ \Young [-1] { 1 1 },  $  etc.)  When full quantum effects are taken into account, only the members of a given
multiplet remain degenerate.    It is a dynamical question which of them eventually become light in the infrared. 

In the softly broken  ${\cal N}=2$  $SU(n_c)$ theory    with $n_f$ flavors, it turns out that the massless non-Abelian monopoles   
appearing in the $r$-vacua   ($q, \, {\tilde q}$)  carry  a flavor quantum number of the fundamental representation   ($\Young{1}$) of the
original flavor $SU(n_f)$  global symmetry group.  This can now be understood as due to  the Jackiw-Rebbi
mechanism  
  generalized to non-Abelian monopoles,  (\ref{possib}).

 The way their condensates break spontaneously the global
$SU(n_f)
\times U(1)$  symmetry  in this theory  is quite interesting.   Upon  ${\cal N}=1$  perturbation  (adjoint mass term  $\mu \Phi^2$),  
condensates  of the form  
\beq      \bra  q_{\alpha}^i   \ket   =  \delta_{\alpha}^i   V, \qquad   \bra  {\tilde q}_{\alpha}^i   \ket   =  \delta_{\alpha}^i   V^{\prime},
\qquad   i, \alpha= 1,2,\ldots, r
\label {ColFl}\eeq 
where $V$, $V^{\prime}$  $\propto  \sqrt { \mu \Lambda },$  develop \cite{CKM}.    On the one hand, they  induce the (dual) Higgs mechanism,
completely  breaking the dual local 
$SU(r)$ group (confinement);  at the same time they trigger  dynamical symmetry breaking
\beq    SU(n_f) \times U(1)  \Longrightarrow    U(r) \times  U(n_f-r).
\eeq
The way the condensate of non-Abelian monopoles break both (dual) color and flavor symmetries   leaving the diagonal subgroup 
unbroken (Color Flavor Locking), shows an
intriguing similarity  to  what is thought  to happen in QCD at high quark densities \cite{CFL}  \footnote{In the softly broken  ${\cal
N}=2$  $SU(n_c)$ theory   under consideration here,   the Yukawa coupling breaks the chiral $SU_L(n_f) \times  SU_R(n_f) $  symmetry to  the
diagonal $SU(n_f)$   at the classical level, so the symmetry breaking pattern  is peculiar to this model.  Another difference is that in the 
high density QCD  it is diquark composites  that condense.   
}.

\section {Duality}

That  the $r$-vacua with  effective, non-Abelian  $SU(r)\times U(1)^{n_c-r}$   gauge symmetry     exist 
only for   $r <  {n_f \o 2}\,$,   thus only  in theories with flavor,        is a manifestation   of the   fact  that  the {\it  quantum  behavior of
non-Abelian monopoles depends crucially on the massless matter fermion degrees of freedom in the fundamental theory}.   
Indeed, the magnetic $SU(r)\times U(1)^{n_c-r}$   theory  with these matter multiplets is  infrared-free (i.e., non asymptotic free).  This is the
correct behavior as it should be  dual to the original asymptotic free $SU(n_c)$  gauge theory.  
Note that  the gauge coupling constant  evolution, which appears as  due to the perturbative loops of magnetic monopoles, 
is actually the result of, and equivalent to,    the infinite  sum of instanton contributions in  the original  $SU(n_c)$   theory.    

This is perfectly analogous to  the observation \cite{KK}  about  how the old paradox related to the   Dirac quantization condition  and
renormali\-z\-ation  group
\cite{Zum} :
\beq     g_e (\mu)  \cdot g_m(\mu) = 2 \pi n, \qquad \forall \mu, 
\eeq   
  is solved within  the 
$SU(2)$  Seiberg-Witten theory. 

This reasoning also leads to the explanation  why in the  pure   ${\cal N}=2$  $SU(n_c)$  theory or on a generic point of the Coulomb
branch of the  ${\cal N}=2$ SQCD,  the
low-energy effective theory is an  Abelian  gauge  theory \cite{SW1}-\cite{curves}.   Massless fermion flavors  
are   needed in order for   non-Abelian monopoles  to  get dressed, via a generalized Jackiw-Rebbi mechanism  discussed in  Section
\ref{JRM}, with  a non trivial  
$SU(n_f)$   flavor quantum numbers  and, as a result,  to render  the dual gauge   interactions  infrared-free. 
When this is not possible,   non-Abelian monopoles are 
strongly coupled and do not manifest themselves as identifiable low-energy degrees of freedom.   

In this respect,   it is very interesting that the boundary   case $r= {n_f \o 2}$   also occurs 
(confining vacua of type (iii) discussed in Introduction)  within the class of supersymmetric theories considered in \cite{CKM}.   In these
vacua,  non-Abelian  monopoles and dyons are strongly coupled, but  still  describes the low-energy dynamics,  albeit via  non-local effective
interactions.     The situation is somewhat similar to what happens in the  ${\cal N}=4$  theories \cite{OlMo}.

\section { Summary and Discussion }

Non-Abelian monopoles are elusive objects.  Though their presence may be detected    in a semi-classical approximation, 
their true nature   depends on the long distance physics.   If the unbroken gauge group  is dynamically  broken
further in the infrared  such multiplets of  states  simply represent an  approximately degenerate  set of magnetic  monopoles.  Only if  
there is no further dynamical breaking do  the non-Abelian monopoles transforming  as nontrivial multiplets of the unbroken,  dual gauge
group,  appear in the theory.

We have shown that there are strong indications that this  occurs   in the $r$-vacua (with an effective $SU(r)\times
U(1)^{n_c-r}$ gauge symmetry) of the softly broken ${\cal N}=2,$    
$SU(n_c)$ supersymmetric QCD \cite{CKM}.  If our
idea  is correct,  this is perhaps the first
 physical  system known  in which Goddard-Nuyts-Olive-Weinberg  monopoles manifest themselves as  infrared degrees of freedom,
playing an essential dynamical role.

More direct verification of the above relation, 
such as  done in  the cases of  Abelian monopoles \cite{KT},     seems to be difficult however, for the subtleties discussed earlier. 

Our arguments apply equally well to the softly broken ${\cal N}=2$ theories with $USp(2n_c)$  or $SO(n_c)$  gauge groups, 
with nonvanishing bare (equal) quark masses.   Semiclassically these gauge groups can be broken to $SU(r)\times U(1)^{n_c-r+1}$, and
non-Abelian  monopoles appearing there can be  identified with the dual quarks  in the low-energy  magnetic $SU(r)\times
U(1)^{n_c-r+1}$ theory.

  Our  observation is   consistent
with  an interpretation suggested in \cite{CKM} (see also \cite{EM}), that these  magnetic quarks might   be  regarded as baryonic constituents
of certain  $U(1)$ monopole, 
\beq     U(1) \, {\hbox{\rm monopole}}   \sim   \epsilon^{a_1   \ldots  a_r} {q_{a_1}^{i_1}  q_{a_2}^{i_2}
\ldots q_{a_r}^{i_r} }.        \eeq
As the  dual $SU(r)$ interactions are infrared-free, 
the Abelian monopole breaks up into its constituents. 
The non-Abelian monopoles  carry a minimum  ${\mathbf Z_r}$  charge:   its  $U(1)$  charge is   ${ 1\o r}$  with respect 
to that of an Abelian monopole (for $U(1)$  lying within the $SU(r)$ subgroup),    in accordance with   the above.  

Another way to relate our massless non-Abelian monopoles to something known,  would be   to consider, within the softly broken
${\cal N}=2$ theories,  the large and equal  bare quark mass regime.  Upon addition of the ${\cal N}=1$ perturbation,  classical
supersymmetric vacua exist with
$SU(r) \times SU(n_c-r)  \\ \times U(1) $  gauge symmetry unbroken and with  $n_f$  massless  quarks  in the ${\underline r} $  of $SU(r)$
\cite{CKM}.  As
this group is infrared free  for $r < { n_f \o 2}$,  the physics at  low energies is  described by the classical Lagrangian.  As $m$ is varied and
reaches values below the dynamical scale
$\Lambda$  of the theory,  however,  the change of monodromy around the quark singularity  occurs, when it  moves below the 
cuts produced by other singularities.  Quarks become magnetic monopoles \cite{SW2}.  Precise way this type of mehamorphosis takes place
has been studied explicitly  only in the simplest  cases of $SU(2)$  theories \cite{CV} (see \cite{Tay}
for some results in $SU(3)$ pure Yang Mills theory)  with Abelian monopoles only.   
Extension of these studies to non-Abelian monopoles seem to be challenging.

  In view of the many known examples of ${\cal N}=1$ Seiberg's duality \cite{seib} in which 
dual magnetic monopoles with non-Abelian charges play prominent roles, 
what is the significance of the present work?   The fact is that in spite of overwhelming indirect evidences that duality features
discussed in \cite{seib} are indeed correct,  the detailed understanding of the magnetic variables appearing in these models  is still lacking. 
For instance, an attempt to ``understand"  the origin of such a duality  starting from the  softly broken ${\cal N}=2$  theory, had only a partial
success
\cite{ArPlSei}.  Any further steps, such as the ones taken here, towards deepening   our insight into the nature of  non-Abelian magnetic
monopoles,   should be welcome.

\section*{Acknowledgments}

The authors  thank    Roberto Auzzi, Roberto Grena,
Jarah Evslin,  Adam Ritz,  Dave  Tong, Alyosha Yung and Valya Zakharov for stimulating  discussions as well as for  useful informations.   One
of the authors (K.K.) thanks the organizers and participants of  the Workshop ``Continuous Advances in QCD",  ArkadyFest,  Minneapolis, May,
2002,  in which some of the ideas in this work  have been presented, for a pleasant atmosphere and for discussions.

\end{document}